Letter

# Resonantly enhanced superconductivity mediated by spinor condensates

Giacomo Bighin,[1] Puneet A. Murthy,[2] Nicolò Defenu,[3] and Tilman Enss[1]

[1]Institut für Theoretische Physik, Universität Heidelberg, Heidelberg, Germany
[2]Institute for Quantum Electronics, ETH Zürich, Zürich, Switzerland
[3]Institute for Theoretical Physics, ETH Zürich, Zürich, Switzerland

(Received 8 March 2024; accepted 30 April 2025; published 20 June 2025)

Achieving strong interactions in fermionic many-body systems is a major theme of research in condensed matter physics. It is well-known that interactions between fermions can be mediated through a bosonic medium, such as a phonon bath or Bose-Einstein condensate (BEC). Here we show that such induced attraction can be resonantly enhanced when the bosonic medium is a two-component spinor BEC. The strongest interaction is achieved by tuning the boson-boson scattering to the quantum critical spinodal point of the BEC where the sound velocity vanishes. The fermion pairing gap and the superconducting critical temperature can thus be dramatically enhanced. We propose two experimental realizations of this scenario, with exciton-polariton systems in two-dimensional semiconductors and ultracold atomic Bose-Fermi mixtures.

Interactions between fermions is the central ingredient in emergent phases, such as superconductivity and the Fractional quantum Hall effect. In conventional superconductors, fermion pairing is mediated by crystal phonons as described by the paradigmatic Bardeen–Cooper–Schrieffer (BCS) theory [1]. Naturally, the strength of the interactions has a direct influence on the critical superconducting temperature ($T_c$) relative to the Fermi temperature ($T_F$). Since phonon-mediated pairing is weak, conventional superconductors exhibit small $T_c/T_F$. This raises the question, Can the properties of the mediator be modified in such a way that enhances the mediated interactions? In this Letter, we propose that the fermionic interactions, and consequently the critical temperature, can be resonantly enhanced by tuning the bosonic medium to a quantum critical point.

In literature, several approaches for enhancing $T_c$ have been explored theoretically and demonstrated experimentally. In twisted bilayer graphene, $T_c/T_F$ can be enhanced by adjusting the twist angle between sheets of graphene to a magic angle of $\approx 1.6^\circ$ [2]. In ultracold atomic ensembles, the fermionic interaction strength can be enhanced by using a Feshbach resonance between two spin components [3]. There have been several theoretical proposals across various experimental platforms for exploiting bosonic mediators such as photons [4,5], exciton-polaritons [6], plasmon polaritons, and atomic Bose-Einstein condensate (BECs) [7–9]. However none of these works harness fine-tuned mediator properties to resonantly enhance the mediated interactions.

In this Letter, we show that coupling a fermionic system to a two-component spinor BEC has profound consequences for induced fermionic interactions. Illustrated in Fig. 1(a), the system comprises orange spheres representing two bosonic species interacting with a two-component Fermi gas (depicted by blue spheres). A key feature of spinor bosonic systems is the existence of a spinodal instability attained through precise tuning of bosonic interactions, where the spin-wave sound velocity vanishes, leading to a quantum critical point. We show that connecting fermions to this quantum critical medium resonantly enhances induced interactions. We analyze these interactions, compute the pairing gap and superconducting temperature, and explore two potential experimental platforms for observation. With respect to similar setups where a single Fermionic species is employed [6,10–12], our study reveals that having two Fermionic species leads to a substantially higher superconductive temperature due to the large pairing coupling in the $s$-wave scattering channel.

*Spinodal instability in two-component BECs.* We examine a 2D system with two fermion species $\uparrow, \downarrow$ embedded in a spinor BEC with two spin components labeled 1 and 2. Initially, we explore key features of the spinor BEC and then address coupled Bose-Fermi mixtures. The spinor BEC has been extensively studied [13–18]. In this context, the transition between miscible and immiscible phases, which occurs when interspecies interactions are large enough [19], has received much attention. Interest in this peculiar phenomenon continues today [20,21].

Coupling two BECs results in new collective normal modes: density and spin waves. Density waves involve in-phase oscillations, while spin waves involve out-of-phase oscillations. The bosonic Hamiltonian, obtained through Bogoliubov transformation, can be diagonalized using density and spin normal modes [13,17,18] and Supplemental Material [22],

$$\hat{\mathcal{H}}^{(b)} = \sum_{\mathbf{q}} \omega^d_{\mathbf{q}} \hat{a}^\dagger_{\mathbf{q}} \hat{a}_{\mathbf{q}} + \sum_{\mathbf{q}} \omega^s_{\mathbf{q}} \hat{b}^\dagger_{\mathbf{q}} \hat{b}_{\mathbf{q}}, \quad (1)$$

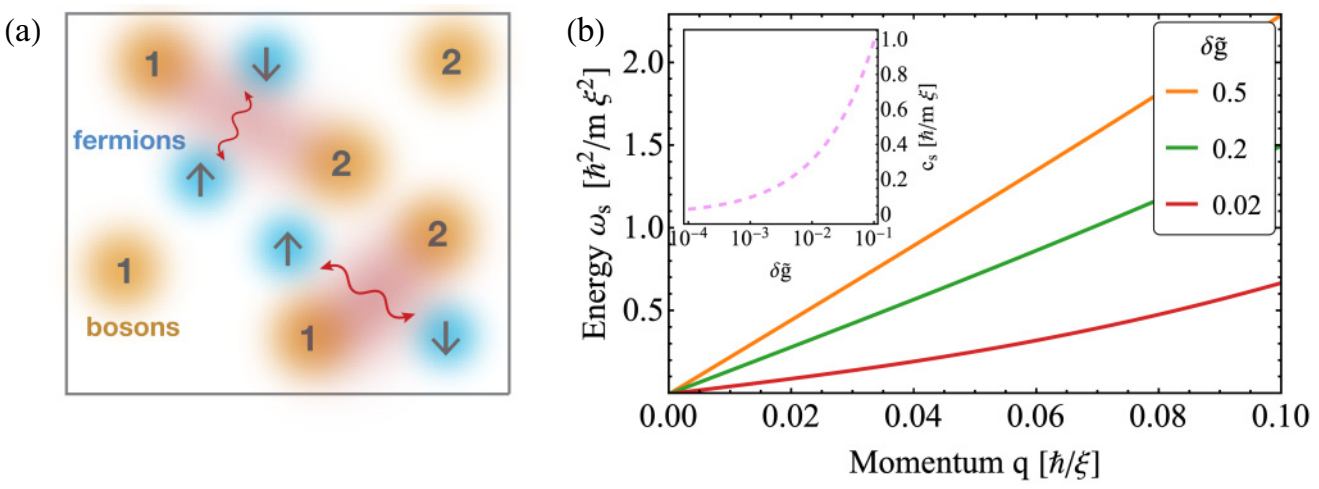

FIG. 1. Fermions in a two-component spinor BEC. (a) Schematic of the induced fermion-fermion pairing mechanism: the spinor BEC normal modes mediate the interaction. By tuning the interboson interaction, one normal mode becomes arbitrarily soft, resonantly enhancing the induced fermion-fermion interaction (red line). (b) Dispersion relation for the spin normal mode of the spinor BEC for three values of detuning $\delta\tilde{g} \equiv (m/\hbar^2)\delta g$ from the spinodal point. Inset: Spin sound velocity $c_s$ softens as $\delta\tilde{g} \to 0$.

where the $\hat{a}^\dagger_{\mathbf{q}}$ ($\hat{b}^\dagger_{\mathbf{q}}$) operators create a density (spin) excitation at momentum $\mathbf{q}$. It is worth noting that Eq. (1) represents the effective Hamiltonian of a generic two-component Bose gas, whatever the origin of the two components, either two internal degrees of freedom, e.g., spins or a Bose gases mixture. In this Letter, we employ the shorthand notation spinor gas to refer generically to all situations where Hamiltonian (1) applies. The Bogoliubov dispersion relation $\omega^{d(s)}_{\mathbf{q}}$ for density (spin) at small momenta reads

$$\omega^{d(s)}_{\mathbf{q}} \approx \sqrt{\frac{\hbar^2 q^2}{2m}\left(\frac{\hbar^2 q^2}{2m} + 2mc^2_{d(s)}\right)} \tag{2}$$

for bosons of mass $m$. The density and spin sound velocities $c_{d(s)}$ depend crucially on the intra- and interspecies interaction strengths between the bosonic components:

$$c^2_{d(s)} = \frac{g_{11}n_1 + g_{22}n_2 \pm \sqrt{(g_{11}n_1 - g_{22}n_2)^2 + 4n_1 n_2 g^2_{12}}}{2m}.$$

Here, $g_{11}$ and $g_{22}$ are the coupling constants between bosons of the same species, while $g_{12}$ is the coupling strength between species 1 and 2. The boson densities of each species are denoted by $n_1$ and $n_2$. The $+$ ($-$) sign corresponds to the density (spin) excitations, respectively.

Careful tuning of the different coupling strengths can lead to dramatic modification of the sound velocities. To see this, we define an effective interaction parameter $\delta g = \sqrt{g_{11}g_{22}} - g_{12}$. For repulsive $g_{12} > 0$ one can tune $\delta g \to 0$, where the spin sound velocity scales as

$$c_s \approx \sqrt{\frac{2g_{12}n_1 n_2}{g_{11}n_1 + g_{22}n_2}}\sqrt{\frac{\delta g}{m}}. \tag{3}$$

An intriguing consequence of Eq. (3) is that for $\delta g \to 0$, the spin sound velocity vanishes even for finite density and scattering lengths. This point is known as the spinodal instability [18], which marks the transition to the immiscible, phase-separated state [23]. Analogously, for $g_{12} < 0$ the miscible state of the spinor BEC exhibits a transition toward a liquid droplet state stabilized by quantum fluctuations [24,25]; near this transition for $\delta g' = \sqrt{g_{11}g_{22}} + g_{12} \to 0$, the spin speed of sound remains finite but instead the density speed of sound becomes soft and vanishes, and with this replacement our conclusions apply equally. In Fig. 1(b), we show the calculated dispersion relation of spin waves for different values of $\delta g$, where the softening of the mode at low momenta can be clearly seen as we approach the spinodal point $\delta g \to 0$ within the miscible phase. In the inset of Fig. 1(b), we show the vanishing of the spin sound velocity $c_s$ as $\delta g \to 0$.

*Induced interactions.* Next, we introduce fermions to the system. The single-particle noninteracting Hamiltonian for fermions has the form

$$\hat{\mathcal{H}}^{(f)} = \sum_{\mathbf{k}\sigma} \epsilon_{\mathbf{k}} \hat{c}^\dagger_{\mathbf{k}\sigma}\hat{c}_{\mathbf{k}\sigma}, \tag{4}$$

where the $\hat{c}^\dagger_{\mathbf{k}\sigma}$ operator creates a fermionic particle of momentum $\mathbf{k}$ and spin $\sigma = \uparrow, \downarrow$. Ultracold Fermi gases can be modeled with a quadratic dispersion relation $\epsilon_{\mathbf{k}} = \hbar^2 k^2/2m_F$ for fermions of mass $m_F$. As the superconducting properties depend predominantly on the density of states $N_0 = m_F/\pi\hbar^2$ near the Fermi surface, our results are robust also for other dispersions $\epsilon_{\mathbf{k}}$ with given $N_0$. For simplicity and to highlight the influence of fermion-boson interactions, we assume that the direct fermion-fermion interactions are negligible; if not, they would be added to our result for the induced interaction [26].

Let us now consider the interaction between fermions and the spinor BEC, modeled by the contact interaction strengths $g_{F1}$ and $g_{F2}$ between either fermion species and boson components 1 and 2, respectively. When written in terms of the normal modes of the spinor BEC, the interaction term in the Hamiltonian reads

$$\hat{\mathcal{H}}^{(f-b)} = \sum_{\mathbf{k};\mathbf{q}\neq 0} M_d(\mathbf{q})\hat{c}^\dagger_{\mathbf{k}+\mathbf{q}}\hat{c}_{\mathbf{k}}(\hat{a}_{\mathbf{q}} + \hat{a}^\dagger_{-\mathbf{q}}) + \sum_{\mathbf{k};\mathbf{q}\neq 0} M_s(\mathbf{q})\hat{c}^\dagger_{\mathbf{k}+\mathbf{q}}\hat{c}_{\mathbf{k}}(\hat{b}_{\mathbf{q}} + \hat{b}^\dagger_{-\mathbf{q}}). \tag{5}$$

Here, $M_d$ and $M_s$ are effective interaction vertices between fermions and the density and spin normal modes of the spinor BEC [27], and we have suppressed the fermion spin index for readability. The interaction vertices are obtained from the original $g_{F1,2}$ after Bogoliubov transformation of the bosons, which couples the fermions to the normal modes of the BEC (see Supplemental Material [22] for details).

In the following, we assume equal boson densities $n_1 = n_2 = n$; our results are qualitatively similar for density or mass imbalanced systems. We find that the effective spin vertex $M_s(\mathbf{q}) \simeq \sqrt{2n}g_F\sqrt{(q^2/2m)/\omega^s_{\mathbf{q}}} \sim \sqrt{q/c_s}$ with $g_F \equiv (g_{F1} - g_{F2})/2$ resembles the conventional electron-phonon coupling introduced by Fröhlich [28]. The same approach also applies to the case of low fermionic density where a polaron state occurs; see Refs. [29–31] for a review. In our case, however, near the spinodal point the spin dispersion in the denominator becomes soft and enhances the interaction vertex toward $M_s(\mathbf{q}) \simeq \sqrt{2n}g_F$, while the density vertex $M_d(\mathbf{q}) \sim \sqrt{q}$ is suppressed for small $q$.

In analogy with the electron-phonon problem, the Bose-Fermi interaction in our system induces interactions between fermions. Following the treatment of Ref. [32], we compute the mediated interaction between two fermions $\mathbf{k}$, $-\mathbf{k}$ that scatter into new states $\mathbf{k}'$, $-\mathbf{k}'$,

$$V_{\mathbf{k}\mathbf{k}'} = -\frac{|M_d(\mathbf{q})|^2\omega^d_{\mathbf{q}}}{\left(\omega^d_{\mathbf{q}}\right)^2 + (\epsilon_{\mathbf{k}'} - \epsilon_{\mathbf{k}})^2} - \frac{|M_s(\mathbf{q})|^2\omega^s_{\mathbf{q}}}{\left(\omega^s_{\mathbf{q}}\right)^2 + (\epsilon_{\mathbf{k}'} - \epsilon_{\mathbf{k}})^2}, \quad (6)$$

where $\mathbf{q} = \mathbf{k}' - \mathbf{k}$ is the exchanged momentum. The exchanged fermionic energy $\omega = \epsilon_{\mathbf{k}'} - \epsilon_{\mathbf{k}}$ captures retardation effects, which become important at strong coupling and tend to lower $T_c$ [12]. The + sign in the denominator arises from renormalization that avoids spurious singularities from normal mode emission and absorption [33]. Near the spinodal point, the contribution from the soft spin modes dominates over the contribution from the density modes, and the induced potential takes the Ornstein–Zernike form

$$V_{\mathbf{k}\mathbf{k}'} \approx -\frac{2ng_F^2q^2/2m}{c_s^2q^2\left(1 + q^2\xi_s^2\right) + (\epsilon_{\mathbf{k}'} - \epsilon_{\mathbf{k}})^2}, \quad (7)$$

with spin healing length $\xi_s = \hbar/2mc_s = \hbar/\sqrt{4m\,\delta g\,n}$. The full two-dimensional momentum dependence of this induced potential is shown in Fig. 2(a): It is strongly enhanced near the Fermi surface and in the forward direction $\mathbf{k} \approx \mathbf{k}'$ ($q \to 0$).

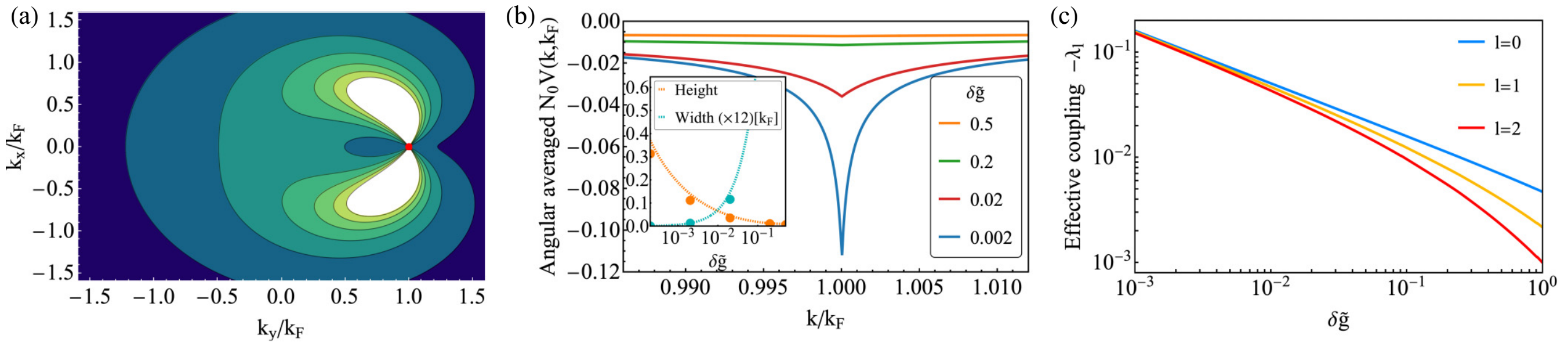

FIG. 2. Effective induced electron-electron interaction. (a) Effective electron-electron interaction $V_{\mathbf{k}\mathbf{k}'}$ in Eq. (7) as a function of $\mathbf{k} = (k_x, k_y)$ near the spinodal instability with $\delta\tilde{g} \equiv (m/\hbar^2)\delta g = 0.001$ and $\tilde{g}_F \equiv (m/\hbar^2)g_F = 0.2$. The second fermion momentum $\mathbf{k}' = (0, k_F)$ is placed at the Fermi surface (red dot). The induced interaction is strongly enhanced in forward direction $\mathbf{k} \approx \mathbf{k}'$. (b) Effective azimuthal-averaged dimensionless interaction $N_0V(k, k' = k_F)$ in the vicinity of the Fermi surface as a function of the radial momentum $k = |\mathbf{k}|$ for different values of $\delta\tilde{g}$, showing the resonant behavior as $\delta\tilde{g} \to 0$. Inset: Height and full width at half maximum of the effective interaction peak at the Fermi surface for various values of $\delta\tilde{g}$, characterizing the resonant scaling behavior as $\delta\tilde{g} \to 0$; the dashed lines guide the eye. (c) Effective coupling $\lambda_\ell$ [Eq. (8)] in angular momentum channels $\ell = 0$ to $\ell = 2$ as a function of $\delta\tilde{g}$.

The enhancement of the induced potential near the Fermi surface is also apparent in the azimuthal average $V(k, k')$ over the angle $\theta$ between the momenta $\mathbf{k}$ and $\mathbf{k}'$. For one fermion $k' = k_F$ fixed at the Fermi surface, the dependence on the other fermion momentum $k$ becomes strongly peaked at $k = k_F$ for smaller values of $\delta g$, as shown in Fig. 2(b). As we approach the spinodal instability $\delta g \to 0$, the peak depth of the induced potential grows as $(\delta g)^{-1/2}$ while the width of the peak is reduced as $(\delta g)^{1/2}$ (see inset).

For scattering between Cooper pairs at the Fermi surface, the fermionic energy exchange in the denominator vanishes, and the induced potential takes the resonance form $V_{\mathbf{k}\mathbf{k}'} \approx -(g_F^2/\delta g)/(1 + q^2\xi_s^2)$. The width of the resonance is given by the inverse spin healing length $\xi_s^{-1}$, which marks the crossover wave number between linear and quadratic spin-wave dispersion. Near the spinodal point $\delta g \to 0$, this becomes much smaller than the standard inverse healing length $\xi^{-1} = \sqrt{2mgn}$ of a single bosonic component. Conversely, the strong induced potential extends to much larger range $\xi_s$ in real space before it crosses over into a van der Waals tail [34].

Since the induced potential at the Fermi surface is strongly peaked in forward direction $\theta \to 0$, a partial wave decomposition of $V_{\mathbf{k}\mathbf{k}'}$ also yields substantial contributions for higher angular momentum channels $\ell = 1, 2, \ldots$ [26]. We define the effective dimensionless coupling $\lambda_\ell$ in the $\ell$ partial-wave channel in two dimensions, in analogy with the three-dimensional case [35],

$$\lambda_\ell = N_0 \int_0^{2\pi} \frac{d\theta}{2\pi}\, \cos(\ell\theta) V_{\mathbf{k}\mathbf{k}'}|_{|\mathbf{k}|=|\mathbf{k}'|=k_F}, \quad (8)$$

with $q^2 = |\mathbf{k} - \mathbf{k}'|^2 = 2k_F^2(1 - \cos\theta)$ for $|\mathbf{k}| = |\mathbf{k}'| = k_F$. The leading couplings for $s$-wave ($\ell = 0$) and $p$-wave ($\ell = 1$) pairing are

$$\lambda_0 = -\alpha\frac{4k_F^2\xi_s^2}{\sqrt{1 + 4k_F^2\xi_s^2}}, \quad \lambda_1 = -\alpha\frac{\left(\sqrt{1 + 4k_F^2\xi_s^2} - 1\right)^2}{\sqrt{1 + 4k_F^2\xi_s^2}}, \quad (9)$$

with dimensionless coefficient $\alpha = (N_0g_F^2/\delta g)/(4k_F^2\xi_s^2) = \frac{1}{2\pi^2}(\tilde{g}_F)^2(mn/m_Fn_F)$, where $n_F = k_F^2/(2\pi)$ denotes the fermion density. The effective couplings $\lambda_\ell$ in different angular momentum channels are shown in Fig. 2(c) as a function of $\delta\tilde{g}$. While far from the spinodal point the $s$-wave channel ($\ell = 0$) dominates, we observe that the $\ell > 0$ contributions become comparable to the $s$-wave channel as we approach the spinodal instability. Near the spinodal instability, $\xi_s \to \infty$ and the interaction in all channels diverges linearly in $\xi_s$ as $\lambda_\ell \approx -\alpha\, 2k_F\xi_s \sim (\delta g)^{-1/2}$ [Fig. 2(c)].

*Pairing gap and critical temperature.* The induced attractive interaction between fermions at the Fermi surface leads to Cooper pairing in analogy with phonon-mediated pairing in conventional BCS superconductors. From the induced interaction potential, we calculate the fermion pairing gap $\Delta(\mathbf{k})$ in our system by solving the BCS gap equation [1,26,36]

$$\Delta(\mathbf{k}) = -\sum_{\mathbf{k}'} V_{\mathbf{k}\mathbf{k}'}\frac{\Delta(\mathbf{k}')}{2E(\mathbf{k}')}\tanh\left(\frac{E(\mathbf{k}')}{2k_BT}\right), \quad (10)$$

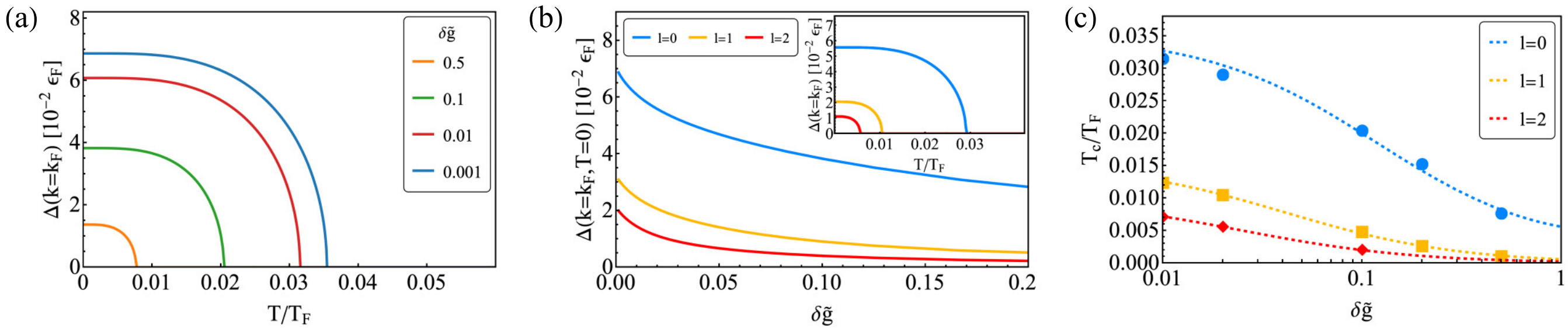

FIG. 3. Fermion-fermion pairing and superconductivity. (a) $s$-wave pairing gap $\Delta_{\ell=0}$ vs temperature $T/T_F$ for different values of detuning $\delta\tilde{g} \equiv (m/\hbar^2)\delta g$, with $\tilde{g}_F \equiv (m/\hbar^2)g_F = 2.0$. (b) Zero-temperature pairing gap $\Delta_\ell$ vs detuning $\delta\tilde{g}$ for different angular momentum channels: All pairing gaps are enhanced for small $\delta\tilde{g}$ near the spinodal point. Inset: Pairing gap $\Delta_\ell$ vs $T/T_F$ for fixed $\delta\tilde{g} = 0.02$. (c) Critical temperature $T_c$ vs detuning $\delta\tilde{g}$; near the spinodal point $T_c$ reaches large values even in higher angular momentum channels because of the resonant induced interaction (dashed lines guide the eye).

with BCS quasiparticle energy $E(\mathbf{k}) = \sqrt{\xi_\mathbf{k}^2 + |\Delta(\mathbf{k})|^2}$ and fermion energy $\xi_\mathbf{k} = \epsilon_\mathbf{k} - \mu_F$ measured from the chemical potential $\mu_F$. At weak coupling, only fermions near the Fermi surface contribute to pairing, and the radial $|\mathbf{k}'|$ integral decouples from the integration over the scattering angle $\theta$ between Cooper pairs at $\mathbf{k}$ and $\mathbf{k}'$. One can then directly use the effective couplings $\lambda_\ell$ from Eq. (8) to obtain the gap $\Delta_\ell \propto \exp(1/\lambda_\ell)$ at zero temperature [26]. At larger coupling, the radial and azimuthal integrals no longer decouple and we find the gap by numerical integration.

In Fig. 3, we characterize the superconducting, paired state by solving the gap equation for different values of the interspecies interactions and for different symmetries of the gap parameter. First, Fig. 3(a) shows the $s$-wave gap $\Delta$, calculated at the Fermi surface as a function of temperature, for different values of the detuning $\delta\tilde{g}$. It clearly shows that pairing is enhanced as the spinodal instability $\delta\tilde{g} \to 0$ is approached and the spin normal mode becomes soft. Second, Fig. 3(b) displays the pairing gap in different angular momentum channels $\ell = 0, 1, 2$. It shows that higher angular momentum channels $\ell > 0$ also have sizable gaps, while the strongest pairing still occurs in the $s$-wave channel ($\ell = 0$). Finally, Fig. 3(c) exhibits the critical temperature $T_c$ in different angular momentum channels. We find that $T_c$, as found from the numerical solution of the gap equation, is significantly enhanced to reach several percent of the Fermi temperature as the resonant regime is approached.

*Experimental realizations.* We now discuss two realistic scenarios where the present theoretical prediction could be experimentally observed: (a) a two-component gas of excitons or exciton-polaritons coexisting with two electron spins in 2D semiconductors and (b) ultracold atomic Bose-Fermi mixtures.

*Exciton-electron systems in 2D semiconductors.* The basic requirement for spinor-BEC mediated superconductivity is a two-component bosonic system with tunable interactions coupled to a fermionic system. This condition is fulfilled in solid-state van der Waals heterostructures consisting of monolayer transition metal dichalcogenide (TMD) semiconductors such as molybdenum diselenide ($MoSe_2$). The elementary bosonic excitations in these systems are excitons, which are optically excited electron-hole bound pairs. The ultrastrong exciton binding energy of about 200 meV in these TMD monolayers means that the excitons are robust and can be considered as pointlike bosons at cryogenic temperatures $T \sim 4$ K [37]. Moreover, the valley degree of freedom in TMD monolayers allows us to selectively address two components (1 and 2) of excitons using circularly polarized illumination. In addition, itinerant charges (electrons or holes) can be introduced in the system through electrical contacts, and their densities can be tuned using proximal gate electrodes. At exciton densities lower than the Mott density ($\sim 10^{13}$ cm$^{-2}$), the interactions between excitons are short-ranged and can be approximated as contactlike. In general, interactions between excitons of the same valley are repulsive with typical values of $g_{11} \simeq g_{22} \simeq 0.1$ μeVμm$^2$ [38,39], whereas the intervalley exciton interactions $g_{12} < 0$ are attractive, leading to a molecular biexciton bound state with binding energy $E_{BX} \simeq 20$ meV.

To control intervalley exciton interactions, various schemes have been proposed [40–42]. In monolayer heterostructures, a Feshbach-like scenario, akin to optical Feshbach resonances in ultracold atomic systems [43], involves a two-photon process [41,42,44]. By detuning an optical field, the intervalley exciton-exciton interaction strength can be tuned from strongly attractive to repulsive [44]. III-V semiconductor heterostructures coupled to optical cavities exhibit a similar scheme with exciton-polaritons [40]. Another approach involves long-lived interlayer exciton condensates in two spatially separated layers, utilizing Feshbach resonance with dc electric fields [45]. In all cases, the goal is to induce critical fluctuations in the two-component Bose gas by tuning intervalley scattering to the spinodal point for repulsive $g_{12} = g = g_{11} = g_{22}$ or to the onset of the droplet state for attractive $g_{12} = -g$.

TMD heterostructures offer the unique ability to inject itinerant electrons (or holes) and tune their density using gate electrodes while maintaining a coexisting population of excitons. Strong attractive interactions between excitons and electrons in different valleys create an intervalley molecular trion-bound state with a binding energy of $E_T \sim 25$ meV [37,46]. Electron densities up to $n_e \sim 10^{12}$cm$^{-2}$ can be achieved by adjusting the gate potential. Resonant illumination allows tuning the exciton density to the same order of magnitude, resulting in a Bose-Fermi mixture with tunable densities and exciton-exciton interactions [38]. Using TMD heterostructure parameters with an effective electron mass $m^* = 0.7m_e$ and $n_e \sim 10^{12}$cm$^{-2}$, the Fermi energy is $\epsilon_F \approx 3.4$ meV, corresponding to a Fermi temperature $T_F \approx 40$ K. Referring to Fig. 3(a) with $\delta\tilde{g} = 10^{-3}$, the estimated $s$-wave zero-temperature pairing gap is $\Delta \approx 0.07\,\epsilon_F \approx 0.24$meV, and the critical superconducting temperature is $T_c \approx 1.3$ K. These values can be experimentally measured in standard cryostats using established four-point dc transport experiments or more advanced ultrafast ac conductivity experiments with THz waveguides if pulsed laser excitation is used to create the exciton gas [47].

*Ultracold atomic Bose–Fermi mixtures.* Mixtures of bosonic and fermionic atoms at ultracold temperatures have been experimentally realized by a number of groups [23,48–52]. As a specific example, we consider a heteronuclear $^{41}$K - $^{87}$Rb Bose-Bose mixture with easily tunable intercomponent scattering length $a_{\text{K-Rb}}$ via a Feshbach resonance at $B = 78.9$G [27,53,54]. The intracomponent scattering lengths are approximately constant at $a_{\text{K-K}} \simeq 62\,a_0$ and $a_{\text{Rb-Rb}} \simeq 100.4\,a_0$ around the Feshbach resonance. This choice is unaffected by spin-changing collisions, ensuring mixture stability, and allows tuning to the spinodal point. Additionally, different hyperfine states of fermionic $^{40}$K interact with the bosonic components via short-range potentials with nonresonant scattering lengths $a_{\text{F-Rb}}$, $a_{\text{F-K}}$ on the order of $50\,a_0$, realizing our proposed scenario of two fermionic components interacting with a Bose-Bose mixture. Predictions for $T_c$ depend on the specific implementation, but as shown in Fig. 3(c) a few percent of $T_F$ can be reached even at moderate detuning from the spinodal point. Various experimental techniques, such as spatially resolved quasiparticle spectroscopy [55,56], matterwave optics for momentum distribution measurement [57], or Bragg spectroscopy [58,59] could be employed to observe the onset of pairing and superfluidity in the fermionic system.

*Discussion.* By fine-tuning bosonic medium interactions, we significantly enhance induced interactions between fermions. Unlike Feshbach resonances, our method achieves resonant interactions while maintaining finite scattering lengths within and between the bosonic species. We achieve controlled spin-wave mode softening near the spinodal point while preserving the phase stiffness of the BEC, ensuring robust collective excitations. This approach fundamentally differs from the instability of a noninteracting BEC with vanishing sound velocity in the presence of fermions [60].

The current system presents a unique form of strongly correlated quantum matter that can test conventional BCS phenomenology to its limits in a real solid-state device. BCS theory has reliably predicted the pairing energy of real fermions throughout the BCS-BEC crossover [56]. As $\delta g \to 0$, the induced interaction surpasses the weak-coupling BCS regime and realizes a solid-state equivalent of the BEC-BCS crossover [3]. Future experimental studies will have to take into account that the location of the spinodal instability can be shifted by interactions beyond the BCS theory, but this will not impact the predicted critical temperature increase. In this perspective, the effective Hamiltonian (1), (4), (5) is to be understood as describing quasiparticles as in Landau Fermi-liquid theory, where the parameter values already include effects due to fermion interactions and mean-field effects due to the bosonic medium [26].

The resemblance between our model and ultracold fermionic atoms extends beyond their BCS description. In both cases, the emergence of strong correlation physics at finite temperature results from an underlying zero-temperature quantum critical point [61]. This quantum critical point separates the homogeneous BEC from a phase-separated or droplet state at $\delta g < 0$ [24,25,62]. The softening of normal mode excitations at criticality substantially increases the effective interaction mediated by the spinor BEC, as shown in Fig. 2(b). This enhanced pairing across all channels is illustrated in Figs. 3(a) and 2(b) and raises critical temperatures as observed in Fig. 3(c).

In conclusion, our predictions describe a way to tailor interactions between Fermi particles that is different from the traditional picture of Feshbach resonances, where one tailors the two-body scattering properties in vacuum to achieve strong many-body correlations. Here, we exploit the presence of a coherent medium, the spinor BEC, whose properties can be altered to control the interaction in the target fermionic system; see Fig. 1. Our framework applies generically and hence can be used to achieve strong correlations both in condensed matter and AMO systems, as we argued in discussing the possible experimental realizations of our proposal. We envisage that the present mechanism could serve to realize a light-induced BCS state whose critical temperature rises well above current limits [63–65].

*Note added.* Recently, two related papers appeared [66,67] that discussed the emergence of superconductivity in two-dimensional TMD heterostructures by mechanisms complementary to ours: The first proposes realizing $p + ip$-wave superconductivity by resonant enhancement of the Bose-Fermi coupling, while the second couples to trions to induce $s$-wave superconductivity. Our proposal pre-dates both of the aforementioned ones.

*Acknowledgments.* We acknowledge stimulating discussions with T. Macrì, A. Imamoğlu, L. Bing Tan, C. Cabrera, L. Tarruell, and K. Fujii at various stages of this work. This work is supported by the Swiss National Science Foundation (SNSF) under Project Funding ID No. 200021_207537, by the Swiss Secretariat for Education, Research and Innovation (SERI), and by the Deutsche Forschungsgemeinschaft (DFG, German Research Foundation) under Germany's Excellence Strategy EXC2181/1-390900948 (the Heidelberg STRUCTURES Excellence Cluster) and under Project-ID No. 273811115 (SFB1225 ISOQUANT).